\documentclass[11pt,a4paper]{article}
\usepackage[utf8]{inputenc}
\usepackage{authblk}
\usepackage{indentfirst}
\usepackage{misccorr}
\usepackage{graphicx}
\usepackage{amsmath}
\usepackage{amssymb}
\usepackage{multicol}
\usepackage{bm}
\usepackage{longtable}
\usepackage{pdflscape}
\usepackage{epstopdf}
\usepackage{multirow}
\usepackage{adjustbox}
\usepackage{amsfonts}
\usepackage{fancyhdr}
\usepackage{ifthen}
\usepackage{fancyheadings}
\usepackage{setspace}
\usepackage[labelfont=bf,font=footnotesize]{caption}

\fancypagestyle{plain}{\chead{\scriptsize \texttt{Communications of BAO, Vol.~2 (LXV), Is.~2, 2018, pp.~???-???}}}
\title{The impact of spiral density waves on the star formation distribution: a view from core-collapse supernovae}
\author{A.~G.~Karapetyan$^{1}$$^{\ast}$, A.~A.~Hakobyan$^{1}$$^{\ast}$, L.~V.~Barkhudaryan$^{1}$,
G.~A.~Mamon$^{2}$, D.~Kunth$^{2}$, V.~Adibekyan$^{3}$, M.~Turatto$^{4}$}
\affil{\emph{\scriptsize$^{1}$Byurakan Astrophysical Observatory, Armenia}\\
\emph{\scriptsize$^{2}$Institut d'Astrophysique de Paris, France}\\
\emph{\scriptsize$^{3}$Instituto de Astrof\'{i}sica e Ci\^{e}ncia do Espa\c{c}o, Universidade do Porto, Portugal}\\
\emph{\scriptsize$^{4}$Osservatorio Astronomico di Padova, Italy}\\
\emph{\scriptsize$^{\ast}$E-mail: karapetyan@bao.sci.am (AGK); hakobyan@bao.sci.am (AAH)}}

\begin{document}
\pagestyle{empty}
\newpage
\pagestyle{fancy}
\label{firstpage}
\date{}
\maketitle

\begin{abstract}
We present an analysis of the impact of spiral density waves (DWs) on
the radial and surface density distributions of core-collapse (CC) supernovae (SNe)
in host galaxies with different arm classes.
For the first time, we show that the corotation radius normalized surface density
distribution of CC SNe (tracers of massive star formation) indicates a dip at corotation
in long-armed grand-design (LGD) galaxies. The high SNe surface density just inside and
outside corotation may be the sign of triggered massive star formation by the DWs.
Our results may support the large-scale shock scenario induced by spiral DWs in LGD galaxies,
which predicts a higher star formation efficiency around the shock fronts,
avoiding the corotation region.
\end{abstract}

\section{Introduction}

According to the pioneering work of Lin \& Shu (1964),
semi-permanent spiral patterns especially in grand-design (GD) galaxies, i.e.
spiral galaxies with prominent and well-defined spiral arms, are created by
long-lived quasi-stationary density waves (DWs).
The results of many observational studies are consistent with the picture where the DWs
generate large-scale shocks and trigger star formation, as originally proposed by Roberts (1969),
causing massive star formation to occur by compressing gas clouds as they pass through
the spiral arms of GD galaxies (e.g. Cepa \& Beckman 1990; Knapen et al. 1996; Seigar \& James 2002;
Grosb{\o}l \& Dottori 2009; Mart{\'{\i}}nez-Garc{\'{\i}}a et al. 2009; Cedr{\'e}s et al. 2013;
Pour-Imani et al. 2016; Shabani et al. 2018).

In this contribution to the international conference \emph{``Instability Phenomena and Evolution of the Universe''}
dedicated to V.~Ambartsumian's $110^{\rm th}$ anniversary (Yerevan-Byurakan, Armenia, 17-21 Sep, 2018),
we briefly present the results of Karapetyan et al. (2018) on the possible impact of spiral DWs
(triggering effect) on the distribution of supernovae (SNe) in discs of host galaxies,
when viewing in the light of different nature of Type~Ia and core-collapse (CC; Types Ibc and II) SNe progenitors,
i.e. less-massive/longer-lived and massive/short-lived stars, respectively.

\begin{figure}
\begin{center}$
\begin{array}{@{\hspace{0mm}}c@{\hspace{0mm}}c@{\hspace{0mm}}}
\includegraphics[width=0.5\hsize]{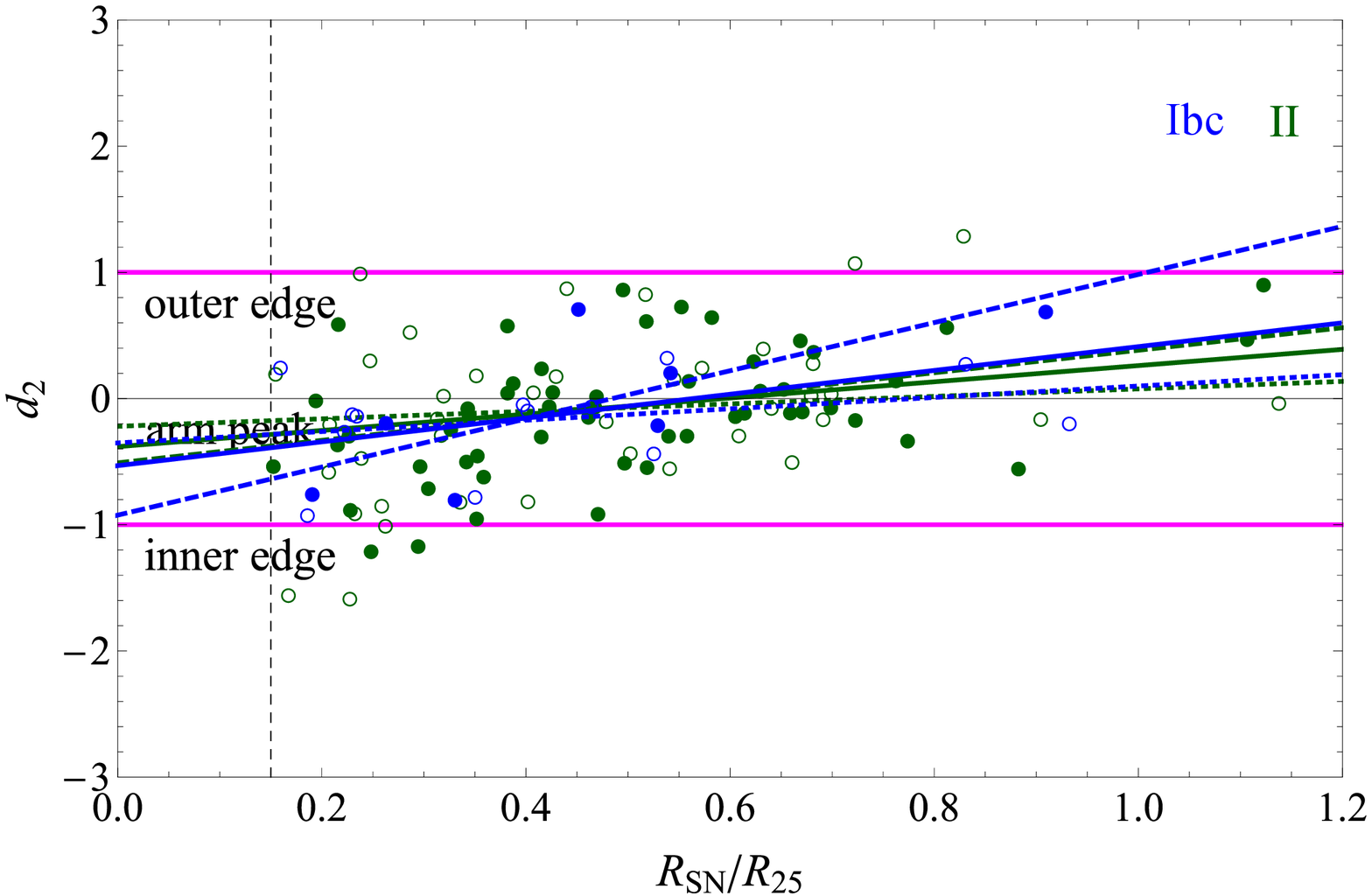} &
\includegraphics[width=0.5\hsize]{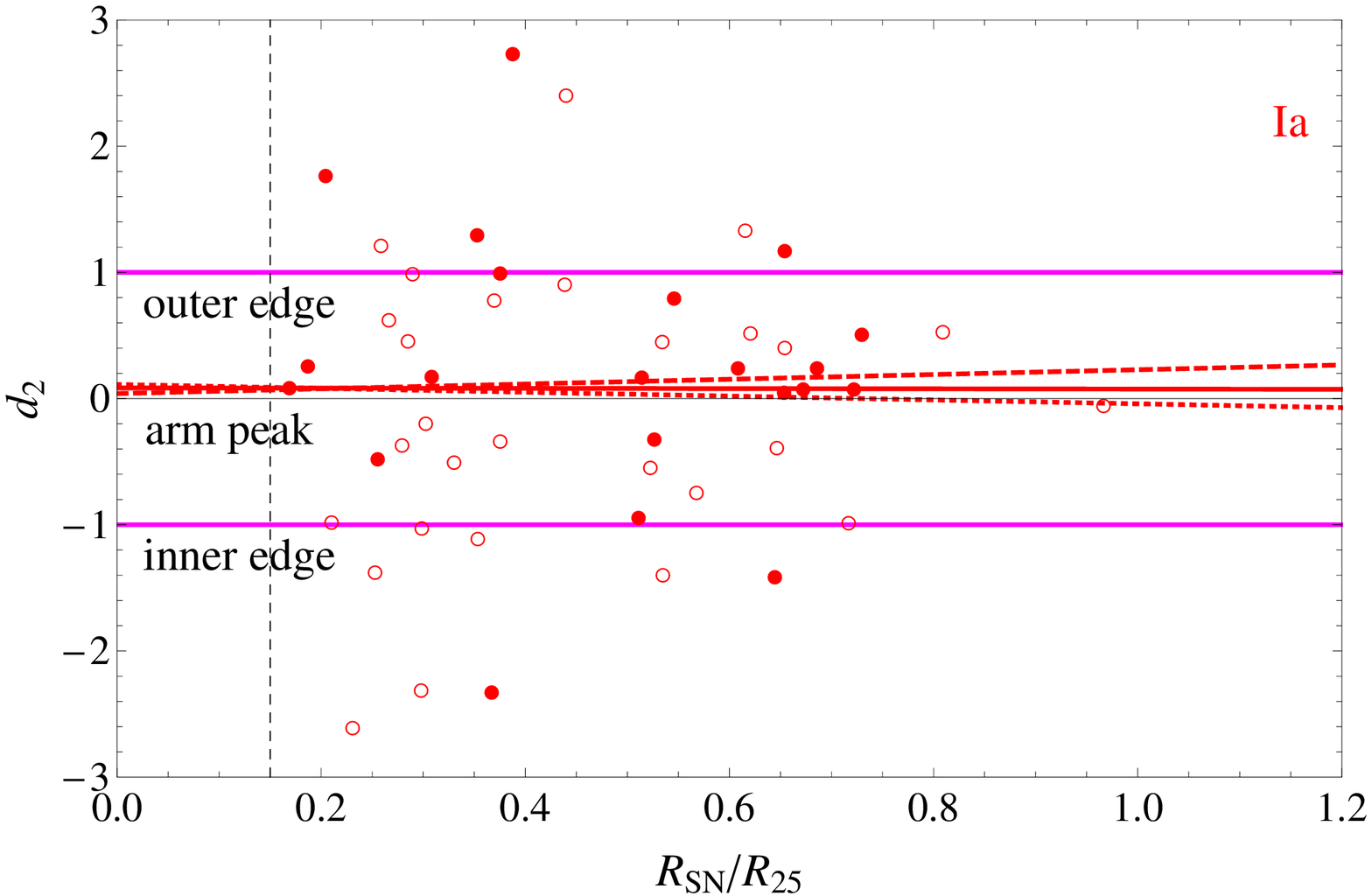}
\end{array}$
\end{center}
\caption{Distributions of the $d_2$ distances of Types Ibc (blue),
II (green) and Ia (red) SNe relative to the peaks of
spiral arms (normalized to the arm semi-width) versus the deprojected and
normalized galactocentric distance ($R_{\rm SN}/R_{\rm 25}$).
Filled and open circles respectively show SNe in GD and NGD galaxies.
In all figures, the linear fits are given
for the full (solid), GD (dashed), and NGD (dotted) samples.
The inner and outer edges, as well the peaks
of spiral arms are shown with parallel lines.
Because of the selection effect near the center of host galaxies,
the region to radius $R_{\rm SN}/R_{\rm 25} = 0.15$
(dashed vertical line) is excluded. This figure is from Aramyan et al (2016),
the reader is referred to the original paper for more details.}
\label{FIG1}
\end{figure}

Although based on small number statistics, the early attempts to study the distribution of CC SNe
within the framework of DW theory were performed by Moore (1973);
McMillan \& Ciardullo (1996) and Mikhailova et al. (2007).
In our recent paper (Aramyan et al. 2016), we
studied the distribution of 215 SNe relative to the spiral arms of their GD
and non-GD (NGD)\footnote{{\footnotesize These galaxies with flocculent arms appear to lack global DWs,
instead their spirals may be sheared self-propagating star formation regions
(see review by Buta 2013, and references therein).}}
host galaxies, using the Sloan Digital Sky Survey (SDSS) images.
We found that the distribution
of CC SNe (i.e. tracers of recent star formation) is affected by the
spiral DWs in their host GD galaxies, being distributed closer to
the corresponding edges of spiral arms where large-scale shocks,
thus star formation triggering, are expected (Fig.~\ref{FIG1}).
Such an effect was not observed for Type~Ia SNe
(less-massive and longer-lived progenitors) in GD galaxies,
as well as for both types of SNe in NGD hosts (Fig.~\ref{FIG1}).

In Karapetyan et al. (2018), we expanded our previous work, and for the first time
checked the triggering effect at different galactocentric radii and studied the
consistency of the surface density distribution of SNe (normalized to the optical
radii [$R_{25}$]\footnote{{\footnotesize The $R_{25}$ is the SDSS $g$-band
$25^{\rm th}$ magnitude isophotal semimajor axis of galaxy.}},
and for a smaller sample also to corotation radii [$R_{\rm C}$] of hosts) with
an exponential profile in unbarred GD and NGD galaxies.

\begin{figure}
\begin{center}$
\begin{array}{@{\hspace{0mm}}c@{\hspace{0mm}}c@{\hspace{0mm}}}
\includegraphics[width=0.51\hsize]{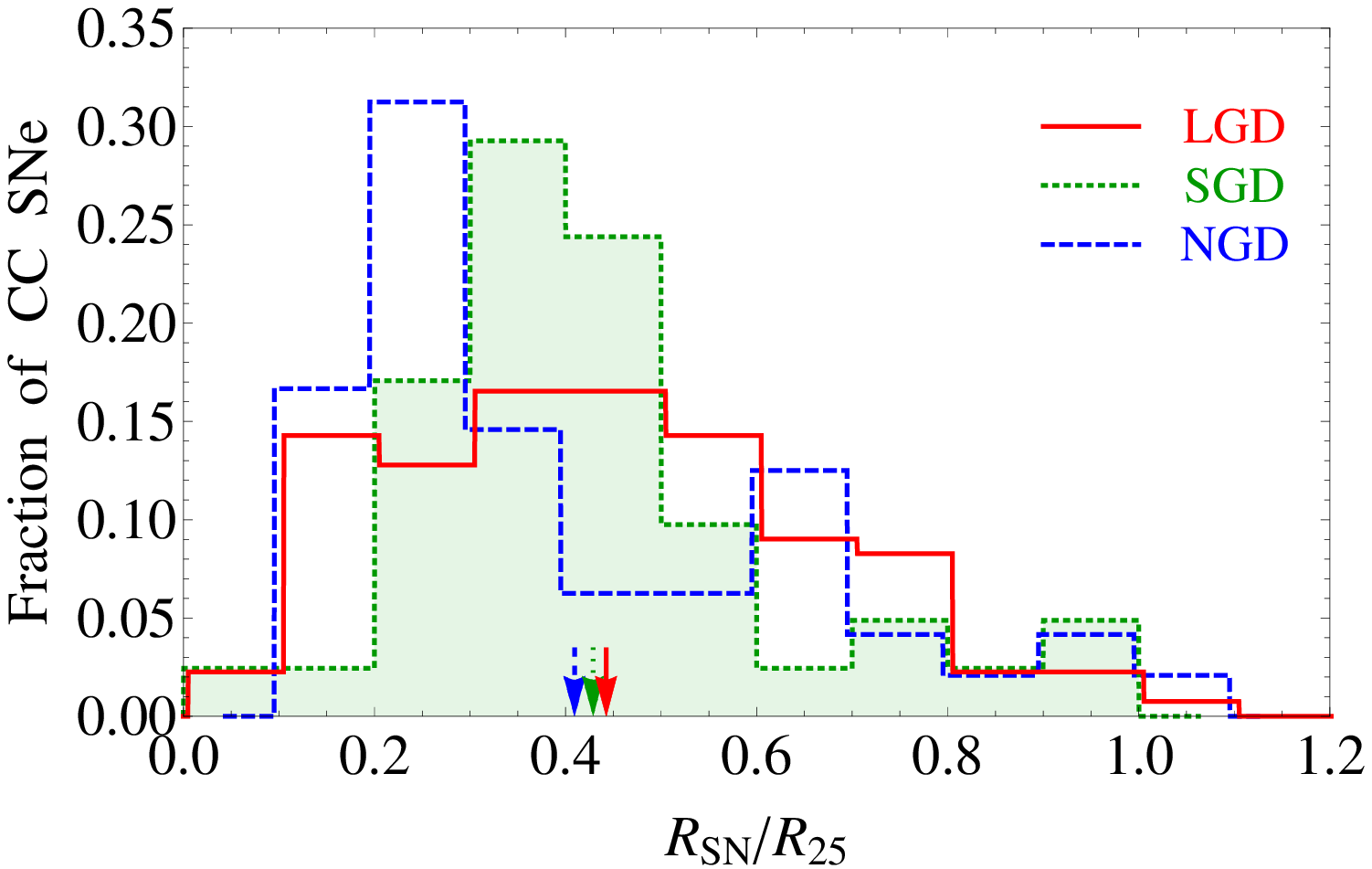} &
\includegraphics[width=0.49\hsize]{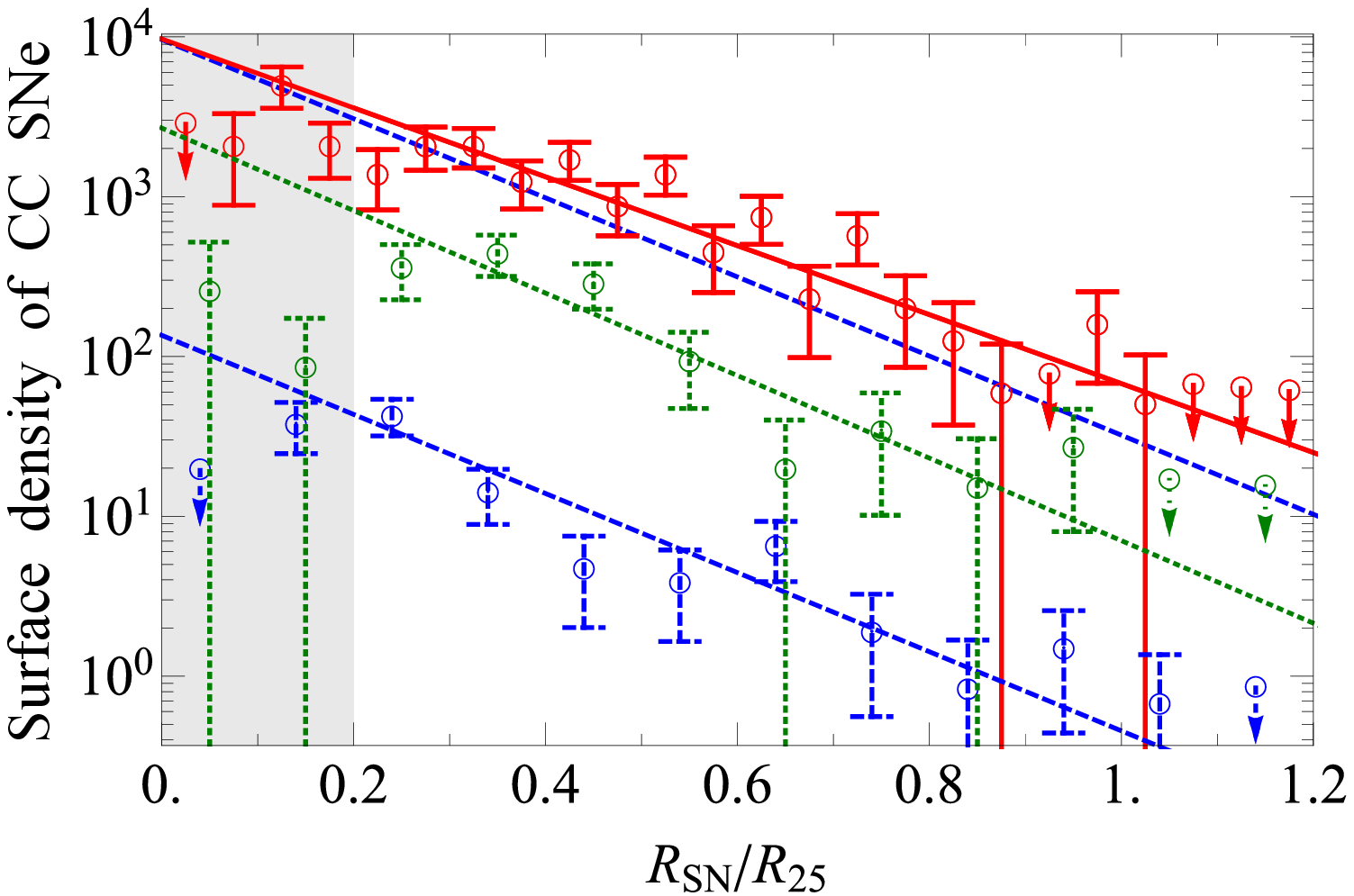}
\end{array}$
\end{center}
\caption{\emph{Left panel}: distributions of deprojected and normalized galactocentric radii ($R_{\rm SN}/R_{25}$) of
CC~SNe in LGD (red solid), SGD (green dotted), and NGD (blue dashed) host galaxies.
The mean values of the distributions are shown by arrows.
\emph{Right panel}: surface density distributions of CC~SNe in the mentioned hosts.
The fitted exponential surface density profiles are estimated for
the inner-truncated discs (outside the shaded area).
For better visibility, the distributions and profiles are shifted vertically sorted by increasing
the mean $R_{\rm SN}/R_{25}$ as one moves upwards, and also slightly shifted horizontally.
To visually compare the distribution of CC~SNe in LGD hosts with the fitted profile in NGD galaxies,
the latter is also positioned with the central surface density
matched with that in LGD hosts. This figure is from Karapetyan et al. (2018),
the reader is referred to the original paper for more details.}
\label{FIG2}
\end{figure}

\section{The sample}

From the coverage of SDSS Data Release~13 (Albareti et al. 2017),
we used a well-defined and homogeneous sample of
SN host galaxies (Hakobyan et al. 2012, 2014, 2016)
with different spiral arm classes according to the classification scheme by
Elmegreen \& Elmegreen (1987).
The sample consists of 269 relatively nearby (${\leq {\rm 150~Mpc}}$),
low-inclination ($i \leq 60^\circ$),
morphologically non-disturbed and unbarred Sa--Sc galaxies, hosting 333 SNe in total.
In addition, we performed an extensive literature search for corotation radii,
collecting data for 30 host galaxies with 56 SNe.
For more details of sample selection and reduction,
the reader is referred to Karapetyan et al. (2018).

\begin{figure}
  \centering
  \includegraphics[width=0.7\hsize]{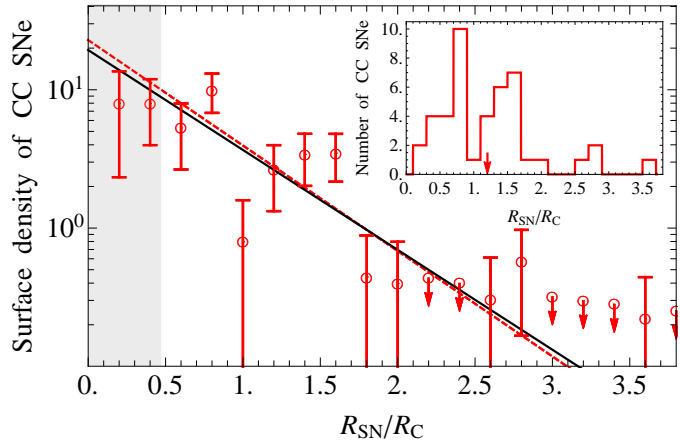}
  \caption{$R_{\rm C}$-normalized surface density profile of CC~SNe
    in LGD host galaxies.
    The black solid and red dashed lines are the best
    maximum-likelihood fits of global and inner-truncated
    (from 0.48 corotation radii outwards to avoid the obscured inner region [grey shaded])
    exponential surface density models, respectively.
    The inset presents the histogram of SN radii (the mean value is shown by arrow).
    This figure is from Karapetyan et al. (2018),
    the reader is referred to the original paper for more details.}
  \label{FIG3}
\end{figure}

\section{Summary of the results and conclusions}

The main results and conclusions concerning the deprojected and
inner-truncated ($R_{\rm SN}/R_{25} \geq 0.2$)
distributions of SNe in host galactic discs are the following:

\begin{enumerate}
\item We found no statistical differences between the pairs of
      the $R_{25}$- normalized radial distributions
      of Type Ia and CC~SNe in discs of host galaxies with different spiral arm classes,
      with only one significant exception:
      CC~SNe in long-armed GD (LGD)\footnote{{\footnotesize The underlying mechanism that
      explains the lengths of arms and their global symmetry in these galaxies is most probably
      a DW, dominating the entire optical disc (e.g. Elmegreen et al. 1992).}}
      and NGD galaxies have significantly different radius distributions.
      The radial distribution of CC~SNe in NGDs is
      concentrated to the centre of galaxies with relatively narrow peak and
      fast exponential decline at the outer region, while the distribution of
      CC~SNe in LGD galaxies has a broader peak, shifted to the outer region of the discs
      (left panel of Fig.~\ref{FIG2}).
      The distribution of SNe in short-armed GD (SGD)\footnote{{\footnotesize The short inner
      symmetric arms in these galaxies might be explained by the DW mechanism,
      dominating only in the inner part of the optical disc (e.g. Elmegreen et al. 1992).}}
      hosts appears to be intermediate between those in NGD and LGD galaxies.
\item The surface density distributions of Type Ia and CC~SNe in most of the subsamples
      are consistent with the exponential profiles (e.g. van den Bergh 1997;
      Wang et al. 2010; Hakobyan et al. 2009, 2016).
      Only the distribution of CC~SNe in LGD galaxies appears to be inconsistent
      with an exponential profile, being marginally higher at
      $0.4\lesssim R_{\rm SN}/R_{25} \lesssim0.7$ (right panel of Fig.~\ref{FIG2}).
      The inconsistency becomes more evident when comparing the same distribution
      with the scaled exponential profile of CC~SNe in NGD galaxies
      (upper blue dashed line in the right panel of Fig.~\ref{FIG2}).
\begin{figure}
  \begin{center}
  \includegraphics[width=0.9\hsize]{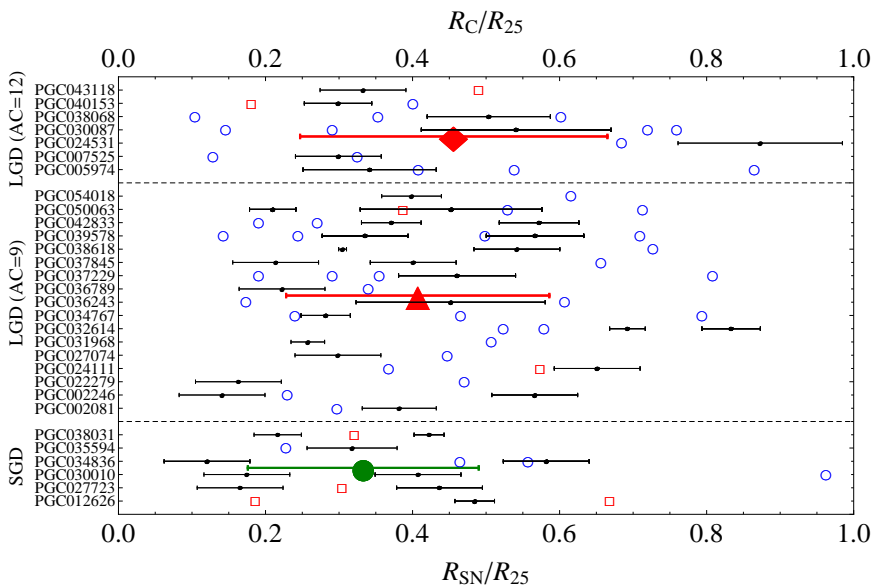}
  \end{center}
  \caption{Galactocentric positions of normalized corotation radii (black points) with their errors
           for SNe host galaxies.
           SGD and LGD galaxies are separated by horizontal dashed lines.
           The spiral arm classes (ACs) of host galaxies correspond to
           the classification by Elmegreen \& Elmegreen (1987).
           The filled diamond, triangle, and circle are the corresponding mean values of
           the corotation radii (with their standard deviations).
           For each host galaxy, galactocentric positions of Type Ia (red empty squares) and
           CC (blue empty circles) SNe are also presented.
           This figure is from Karapetyan et al. (2018),
           the reader is referred to the original paper for more details.}
  \label{FIG4}
\end{figure}
\item Using a smaller sample of LGD galaxies with estimated corotation radii,
      we showed, for the first time, that the surface density distribution of
      CC~SNe shows a dip at corotation, and enhancements at $^{+0.5}_{-0.2}$
      corotation radii around it (Fig.~\ref{FIG3}).
      However, these features are not statistically significant.
      The CC~SNe enhancements around corotation may, if confirmed with larger
      samples, indicate that  massive star formation is triggered by the DWs in LGD host galaxies.
      Considering that the different LGD host galaxies have various corotation radii
      (Fig.~\ref{FIG4}) distributed around the mean value of
      $\langle R_{\rm C}/R_{25}\rangle=0.42\pm0.18$,
      the radii of triggered star formation by DWs are most probably blurred
      within a radial region including $\sim0.4$ to $\sim0.7$ range in units of $R_{25}$,
      without a prominent drop in the mean corotation region (right panel of Fig.~\ref{FIG2}).
\end{enumerate}

These results for CC~SNe in LGD galaxies may, if confirmed with larger samples
and better corotation estimates, support the large-scale shock scenario
(e.g. Moore 1973), originally proposed by Roberts (1969),
which predicts a higher star formation efficiency, avoiding the corotation region
(e.g. Cepa \& Beckman 1990; Seigar \& James 2002; Cedr{\'e}s et al. 2013; Aramyan et al. 2016).

When more information will become available on corotation radii of SN host galaxies,
it would be worthwhile to extend our study, by comparing the $R_{\rm C}$-normalized
radial and surface density distributions of Type Ia and CC~SNe in LGD galaxies.
This will also allow to check the impact of spiral DWs
on the distribution of less-massive and longer-lived progenitors of Type Ia SNe.
Moreover, similar analysis of SNe in SGD galaxies can help to understand the role of DWs
in star formation triggering, if any.

\section*{\small Acknowledgements}

\scriptsize{This work was supported by the RA MES State Committee of Science, in the
frames of the research project number 15T-1C129.
This work was made possible in part by a research grant from the Armenian
National Science and Education Fund (ANSEF) based in New York, USA.}

\end{document}